# Composite Fixed-Length Ordered Features for Palmprint Template Protection with Diminished Performance Loss

Weiqiang Zhao[1], Heng Zhao[1], Zhicheng Cao[1], and Liaojun Pang[1*]

[1] School of Life Science and Technology, Engineering Research Center of Molecular and Neuro Imaging, Ministry of Education, Xidian University, Xi'an 710126,
Shaanxi, China

## Abstract

Palmprint recognition has become more and more popular due to its advantages over other biometric modalities such as fingerprint, in that it is larger in area, richer in information and able to work at a distance. However, the issue of palmprint privacy and security (especially palmprint template protection) remains under-studied. Among the very few research works, most of them only use the directional and orientation features of the palmprint with transformation processing, yielding unsatisfactory protection and identification performance. Thus, this paper proposes a palmprint template protection-oriented operator that has a fixed length and is ordered in nature, by fusing point features and orientation features. Firstly, double orientations are extracted with more accuracy based on MFRAT. Then key points of SURF are extracted and converted to be fixed-length and ordered features. Finally, composite features that fuse up the double orientations and SURF points are transformed using the irreversible transformation of IOM to generate the revocable palmprint template. Experiments show that the EER after irreversible transformation on the PolyU and CASIA databases are 0.17% and 0.19% respectively, and the absolute precision loss is 0.08% and 0.07%, respectively, which proves the advantage of our method.

---

[*] Send emails to corresponding author for further questions: ljpang@mail.xidian.edu.cn.

# I. Introduction

Compared with passwords or ID cards, the identity authentication means of biometrics is more convenient, unique and permanent, and can effectively combine with physical and digital identities [1,2]. As a more and more popular biometric modality, the palmprint has richer feature information than fingerprint, face, iris, etc., and the acquisition equipment is cheap and convenient. Therefore, palmprint recognition has been widely applied in high security scenarios such as ATM and safe box [3].

However, images and templates may be leaked during the process of palmprint recognition. Because users cannot directly revoke and re-publish their palmprint templates, this results in serious data storage insecurity and privacy breakage [4]. An effective way to solve this problem is the biometric template protection (BTP) technology [5], which processes the original palmprint features using a certain template generation algorithm, and stores the generated new templates in the system instead of the original palmprint features.

A good palmprint template protection scheme should have four characteristics of biometric template protection: authentication performance, diversity, revocability and irreversibility[6]. Among them, the authentication performance refers to the transformation of the template after authentication performance, the diversity refers to that the transformed template in different databases must be sufficiently different, and the revocability refers to the system is possible to republish a new transformed template with sufficient differences from the previous template when the transformed template is leaked or lost, and the irreversibility refers to the information that the original palmprint template cannot be inferred from the transformed template. And diversity, revocability and irreversibility are all important components of security in palmprint template protection schemes.

So far, the research work on palmprint template protection is still insufficient, though some progress has been made. Current palmprint template protection methods are either based on biological hashing or irreversible transformation, both of which are still far from ideal in the performance. Most methods of irreversible transformation only use the directional features of the palmprint with transformation processing, yielding unsatisfactory protection and identification performance.

From the perspective of improving palmprint feature extraction, this paper proposes a palmprint template protection method based on the fusion of directional and point feature. On one hand, in order to solve the problem of inaccurate directional feature extraction, a new direction feature extraction algorithm based on MFRAT of two directions fusion is proposed to obtain more accurate palmprint direction feature; On the other hand, in order to solve the problem of insufficient utilization of palmprint features, through block processing and SURF [7] algorithm to get fixed-length ordered point feature, IOM [8] irreversible transformation protection template is used after fusing point features and direction features, the fusion of the two features further enriches the feature information of the extracted palmprint and improves the authentication accuracy.

The main contributions of this paper are:

1. This paper proposes a new directional feature extraction method based on MFRAT of two directions fusion. The existing palmprint direction feature extraction methods in the research of palmprint template protection are not accurate enough, for a pixel point on a palmprint image, the main direction feature of the point is obtained by using the MFRAT template, then the relationship between the main direction and the adjacent direction is used to obtain a more accurate directional characteristic of the pixel.

2. This paper proposes a new fixed-length ordered point feature extraction method. Because the point features are usually indefinite and disorderly, it is difficult to directly use for feature fusion and irreversible transformation, in this paper, the palmprint image is divided into blocks, and the SURF algorithm is used to obtain the most representative feature points for each block to obtain fixed-length ordered point features, laid the foundation for subsequent feature fusion and irreversible transformation processing.

3. This paper fuses point feature and direction feature, and the fused feature is

irreversibly transformed to generate a revocable palmprint template. At present, most palmprint template protection methods only use directional features for processing, and the authentication accuracy is always difficult to improve, by using the feature fusion strategy, the authentication performance of the proposed scheme in this paper has been further improved compared with other schemes using a single feature. In addition, in the research of palmprint template protection, on the basis of obtaining the fixed-length ordered fusion feature, this paper uses IOM irreversible transformation for the first time to generate a revocable palmprint template.

The rest of the paper is organized as follows: the second part is mainly about the related work, the third part is mainly the detailed description of the proposed method, the fourth part is the experimental design and experimental results, and the last part is the summary and prospect.

## II. Related Works

Biometric template protection according to the description of Jain [9] and others can be roughly divided into two categories: biometric transformation and biometric encryption. Biometric encryption technology mainly includes key binding technology represented by fuzzy commitment [10] and fuzzy safe [11], and key generation technology represented by security abstract fuzzy extractor [12], combining biometrics with cryptography, although some research results have been obtained, due to the contradiction between the sensitivity of the key and the differences within the biometric class makes the authentication performance difficult to reach the ideal standard. At present, most researches on palmprint template protection are based on feature transformation [13], the basic idea is to transfer the original biological features to another domain using transformation function, generate and store the security template according to the transformed information, and finally complete the template matching in the encryption domain. For the palmprint template protection method based on feature transformation, selecting the appropriate palmprint features for extraction and appropriate transformation function design are the keys to ensuring security and authentication performance.

According to the characteristics of the transformation function, palmprint template protection based on feature transformation is generally divided into two categories: based on the biological hash (BioHash) method and the irreversible transformation method. For the BioHash method, its original idea was proposed by Teho et al.[14], which defined an orthogonal random transformation function through a specific key, and used the function to transform the biological features to obtain the hash sequence. Based on the BioHash algorithm, Leng et al.[15] proposed a PalmHash palmprint template protection algorithm in combination with palmprint features, and the algorithm is also diverse, revocable, irreversible, and highly secure, but because the algorithm directly binarizes the Gabor filter result and the inner product of the random matrix to obtain the PalmHash Code, its authentication performance is average and the EER is 0.8%. In order to further improve security and authentication performance, Leng[16] et al. proposed a PalmPhasor palmprint template protection algorithm based on BioPhasor[17] algorithm. Compared with PalmHash algorithm, it uses non-linear operation instead of linear operation, on the one hand, the irreversibility of the algorithm was improved to ensure the security of the template, on the other hand, the authentication performance has greatly improvement, and the EER is 0.4%. Although the above palmprint protection algorithm based on BioHash has made some breakthroughs in security and authentication performance, the BioHash method relies heavily on the security of user passwords. When a user's password is lost or leaked, the attacker will impersonate a real user with a combination of the user's password and biometrics to perform identity authentication, resulting in greatly reduced authentication performance of such methods.

In order to avoid the problem of reduced authentication performance caused by the loss of user passwords in the BioHash method, researchers began to use irreversible transformation functions to implement template protection for palmprints. The palmprint template protection method based on irreversible transformation mainly performs irreversible transformation on palmprint features, so that the transformed features cannot recover the original features, thereby ensuring the security of palmprint features. This method was originally proposed by Ratha[18] et al., and was first applied to the protection of fingerprint feature templates, its main idea is to map fingerprint minutia features into new features through three methods: position transformation, polar coordinates and function transformation, so as to store and compare the transformed template. Feng[19] et al. analyzed the three transformation methods designed by Ratha, and considered that these transformation functions were relatively insecure and could reconstruct original biological features through a large number of attacks. Based on this, Li[20] et al. first implemented a chaotic high-speed stream cipher algorithm based on the coupled nonlinear dynamic filter (CNDF) to generate the updatable and privacy-protected palm print template. This scheme uses multiple sets of Gabor filters in different directions to extract the phase features of the palmprint in the feature extraction phase, so that the generated revocable palm print template has greater inter-class differences, the security and the authentication performance are improved. Similarly, using the direction feature, Qiu et al.[21] proposed a new revocable palmprint template generation scheme based on a lookup table. The lookup table method used in this scheme is relatively simple and has the advantage of fast speed, the final authentication performance is hardly satisfactory. To further improve the security of palmprint feature templates, Qiu et al. [22] proposed a palmprint template protection scheme based on random comparison and noise data. The advantage of this scheme is that even in the worst case of the password theft, a high level of security can be achieved by calculating the proportion that adjust noise data reasonably. In addition, Kaur et al. successively proposed a palmprint template protection scheme based on random slope [23] and random distance [24], which further solved the problem of template protection when user passwords were leaked. Compared with the BioHash method, the palmprint template protection method based on irreversible transformation has made great progress in security, and the authentication performance has also been improved.

In summary, in the current research on palmprint template protection, a more secure, reliable, and high-precision revocable palmprint template is mainly generated by reasonably extracting palmprint features and selecting a proper transformation function. At present, most of the palmprint template protection methods mainly use the directional features of palmprint to generate the template through irreversible transformation, although notable breakthroughs have been made in diversity, reversibility and irreversibility, the final authentication performance is still not ideal. Since different palmprint features have their own advantages and disadvantages, in order to further improve the authentication performance, we can consider introducing other palmprint features besides directional features for template protection. The point feature is also an important feature in palmprint, by using the point features and information around them, palmprint images can be accurately identified and protected by the template. Compared to using a single texture direction feature, the extraction and fusion of a variety of palmprint features can make full use of the inherent feature information of palmprint, and further improve the authentication performance of palmprint template protection.

## III. The Proposed Method

This part details the proposed palmprint revocable template generation method based on the fusion of point feature and directional feature. This method mainly consists of three parts, as shown in Figure 1, they are feature extraction based on MFRAT of two directions fusion, feature extraction based on SURF algorithm for fixed-length ordered points, and generated revocable palmprint template by using the IOM irreversible transformation after the point feature and the direction feature fusion.

### A. Feature extraction based on MFRAT of two directions fusion

Direction features are common features in palmprint images, Robust Line Orientation Code (RLOC)[25] based on MFRAT template filtering is a more commonly used method for direction feature extraction. Although this method can quickly detect the direction features of palmprints, the extracted direction features are not accurate enough. We made experimental statistics on the deviation of the direction feature extracted by RLOC algorithm in PolyU database, and found that the minimum response direction of 82.6% of pixels is adjacent to the next-smallest response direction, 21.4% of these pixels have their true primary direction falling between the minimum response direction and the second-minimum response direction. A specific example is shown in figure 2, the response value of the pixel in the middle direction between the minimum response direction and the second-minimum response direction is smaller than the minimum response value, indicating that the true primary direction should be between the minimum response direction and the second-minimum response direction.

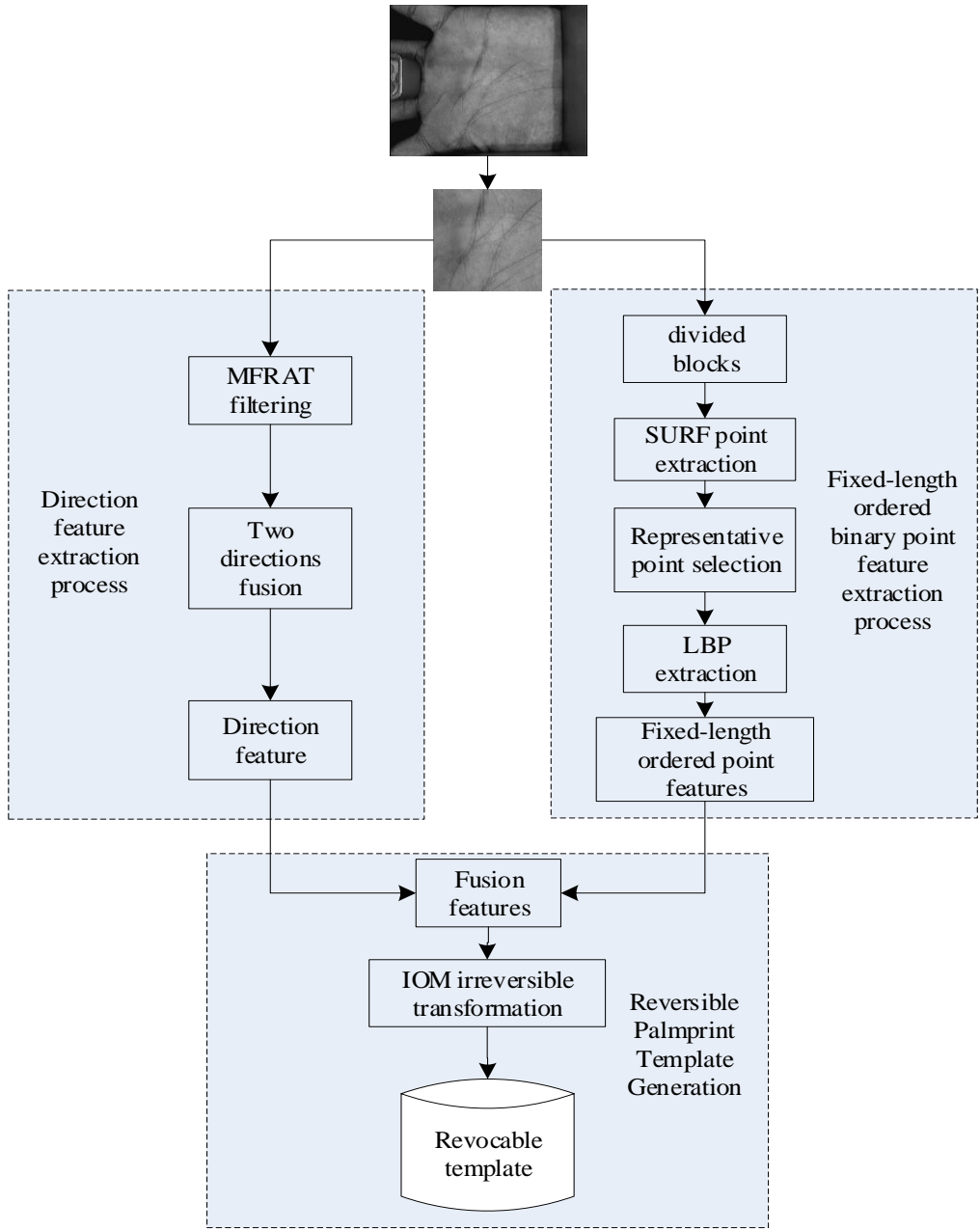

Fig. 1. The algorithm of generating reversible palmprint template

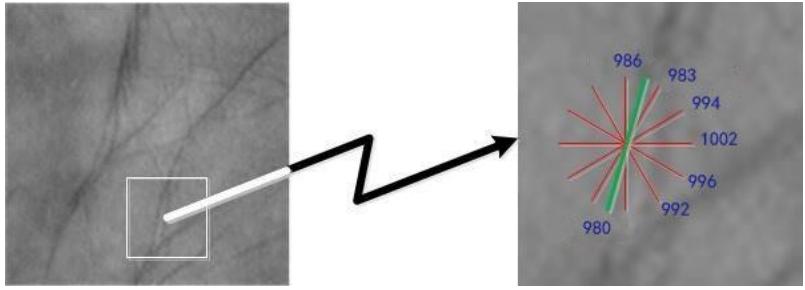

Fig. 2. Comparison of MFRAT response values: the green line represents the minimum MFRAT response value of the pixel under actual conditions, and the red line represents the response value obtained by using MFRAT templates in 6 different directions.

In order to further improve the detection accuracy of direction features, this paper proposed a feature extraction algorithm based on MFRAT of two directions fusion, which judges and fuses the direction features obtained from MFRAT templates in 6 different directions, and extract more accurate palmprint direction features without increasing the MFRAT template. When the minimum response direction of a pixel value in the palmprint image is adjacent to the second-minimum response direction, and the difference between the second-minimum response value and the minimum response value is less than the preset threshold, the true main direction of the pixel should be the middle between the minimum response direction and the second-minimum response direction. Based on this, we use the correlation between the minimum response and the next smallest response to extend the direction retrieval range to 12 directions based on the MFRAT template that still uses 6 different directions, as shown in Figure 3. Assume that the directions of the 6 MFRAT templates used at the pixels are $(\pi/6)q(q = 0,1,...,5)$, then the 12 directions obtained by the expansion of two directions fusion are $(\pi/12)o(o = 0,1,...,11)$, the specific calculation formula is as follows:

$$o(x,y) = \begin{cases} 2q_{min} & |mod(q_{min} - q_{sec}), 5| == 1 \text{ and } f_{sec} - f_{min} < r \\ q_{min} + q_{sec}, & else \end{cases} \quad (1)$$

Among them, $q_{min}$ represents the direction corresponding to the minimum MFRAT response at the pixel $(x,y)$, $q_{sec}$ represents the direction corresponding to the

MFRAT second-minimum response of pixel$(x,y)$, $f_{min}$ represents the minimum response value of MFRAT, $f_{sec}$ is the MFRAT second-minimum response value, $o(x,y)$ is the direction obtained after MFRAT of two directions fusion, $|mod(q_{min} - q_{sec}), 5| == 1$ means that the minimum response direction and the second-minimum response direction are adjacent. When the minimum response direction is adjacent to the second-minimum response direction and the difference between the minimum response value and the second-minimum response value is less than the set threshold $r$, at this time, $f_{min}$ is not the actual minimum response value, the direction $o(x,y)$ obtained by the two directions fusion should fall between the minimum response direction and the second-minimum response direction, which is $q_{min} + q_{sec}$.

Through MFRAT of two directions fusion, the direction retrieval range is extended from [0,5] to [0,11] without increasing the MFRAT template, which improves the accuracy of palmprint direction feature coding. Compared with the traditional RLOC algorithm, the proposed algorithm based on MFRAT of two directions fusion in this paper makes good use of the correlation between the minimum response and the second-minimum response, and obtains more accurate directional features without increasing the amount of calculation.

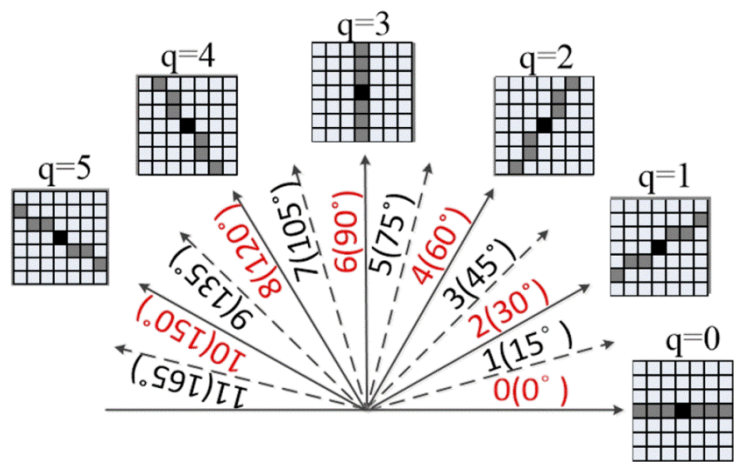

Fig.3. MFRAT of two directions fusion

**B. Feature extraction of fixed-length ordered points based on SURF algorithm**

Point feature is an important feature in palmprint, the point feature extracted by

SUFR algorithm has the advantages of scale invariance, rotation and translation invariance and affine invariance. Because the point features obtained by directly using the SURF algorithm in palm print images are of variable length and out of order, it is difficult to be used for feature fusion and revocable template generation. In the matching process, the features of fixed-length ordered points are more convenient and faster than those of indefinite-length disordered points, Figure 4 shows the matching process of the above two kinds of point features, in which (a) is the matching process of indefinite-length unordered point features, and (b) is the matching process of fixed-length ordered point features. In addition, compared with the fixed-length ordered feature, the indefinite-length unordered feature is more difficult to match in the encryption domain after irreversible transformation, so we need to perform fixed-length ordered processing on indefinite-length unordered point features.

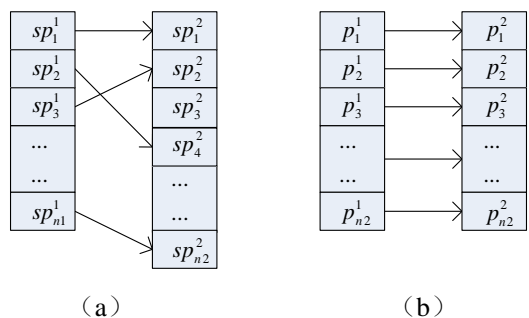

Fig.4. Schematic diagram of matching of two kinds of point features

In this part, we obtained the most representative SURF feature points in each block by using SURF algorithm for palmprint image blocks, and finally obtained the fixed-length ordered point feature by using LBP operator [26] for each representative point. The specific steps are as follows:

1) The original palm print image is divided into non-overlapping blocks, and the edges are complemented to obtain $m$ palmprint block images $I_i$ of size $n \times n$. The image block was processed by SURF algorithm to determine whether the pixel point $I_i(x, y)$ was a SURF point, and record the detected SURF point $sp_i$.

2) The minimum distance method was used to select the most representative

SURF feature points in the block to solve the problem that multiple SURF points are close to each other exist in some block images. Suppose there are $n_i$ SURF feature points in the $i$-th image block, and the sum of the distances between the $j$-th feature point $sp_i^j$ and other feature points in this image block is $d_i^j$, then we can get the distance from other SURF feature points and the minimum feature point index value $D_i$:

$$D_i = argmin\{d_i^1, d_i^2, \ldots, d_i^{n_i}\} \quad (2)$$

Then the $D_i$ SURF feature point is the most representative SURF feature point in the image block, and it is named as the representative point $RP_i$. Due to select the unique representative point in each block image, the number of representative points of the entire palmprint image is determined by the number of blocks, based on this, the number of representative points extracted for different palm print images is the same, in addition, because the processing is performed block by block, the extraction order of representative points is fixed, so the final representative points are ordered and the number is fixed.

3) Using the representative points $RP_i$ and the LBP operator to obtain a coded value $p_i$ as the point feature of the image block, the point feature of the entire image is finally obtained as $P = [p_1, p_1, \ldots, p_m]$. The order and number of representative points extracted from different palmprint images are the same, which indicates that the point features we finally extract are fixed-length and ordered.

After processing the image block, the pseudo code of the fixed-length ordered point feature extraction algorithm is as follows:

| **Algorithm 1** Feature extraction of fixed-length ordered points based on SURF algorithm |
|---|

**Input**   block image $I_i$, size of block image is $n \times n$, number of block images is $m$

**Stage 1:** SURF feature points were determined one by one image block

   if   $I_i(x,y)$  is the SURF feature point, then

$sp_i \leftarrow (x,y)$

   end if

   Save $sp_i, n_i$      //$sp_i$  and  $n_i$ are the SURF feature points and their number of the *i-th* image block

**Stage 2:** For each image block, select the representative point $RP_i$ according to the sum of the distance $d_i$ between each SURF point and other SURF feature points

$D_i = \text{argmin}\{d_i^1, d_i^2, \ldots, d_i^{n_i}\}$

$RP_i = sp_i(D_i)$

   Save  $RP_i$      //$RP_i$ is the representative point of the *i-th* image block

**Stage 3:** Process the representative point $RP_i$ one by one to obtain the fixed-length ordered point feature $P$

$p_i = \sum_{L=0}^{L-1} 2^L s(I_L - RP_i), \quad L = 0,1,\ldots,7$      // use LBP operator centered on $RP_i$

$s(x) = \begin{cases} 1 & if\ x \geq 0 \\ 0 & else \end{cases}$

$P = [p_1, p, \ldots, p_m]$

**Output**   fixed-length ordered point feature  $P$

### C. Feature fusion and revocable template generation

In biometric recognition, fusion is a common strategy, and the accuracy of matching can be improved through fusion. Fusion is generally divided into feature layer fusion, fractional level fusion and decision-level fusion, in this paper, since the extracted directional feature $O$ and point feature $P$ are both fixed-length ordered feature coding values, fusion can be conducted directly at the feature layer to obtain the fused feature $C$:

$$C = [O\ P] \qquad (3)$$

After the fused features are obtained, they need to be irreversibly transformed to generate a reversible palm print template. Index-of-Max (IOM) irreversible transformation algorithm achieves excellent authentication performance on the basis of revocation, irreversibility, and diversity. The basic idea is to non-linearly embed fixed-length ordered fingerprint minutia into a rank metric space based on the similarity function, and generate a revocable fingerprint template by recording the index of the maximum projection feature value. In this paper, the IOM is first applied in the palmprint template protection work, because the direction feature and point feature of the proposed approach are fixed-length and ordered, the features after fusion are also fixed-length and ordered, therefore, IOM can be used to perform irreversible transformation processing to generate security and revocable palmprint template, the specific reversibility analysis can be found in [8]. The specific steps to generate the revocable template based on IOM irreversible transformation are as follows:

1) Generate $l$ $k$-dimensional random gaussian projection vectors to build a random Gaussian projection matrix $W^i$:

$$W^i = [w_1^i, w_2^i, ..., w_k^i] \qquad (4)$$

where, $i = 1,2,...,l, j = 1,2,...,k.$

2) Multiply the fused feature with a random gaussian projection vector and record the index $X_i$ of the maximum value multiplied:

$$X_i = argmax_{j=1,2,...,k}\langle w_j^i, H \rangle \qquad (5)$$

3) Get the revocable palmprint template $X = \{X_i \in [1,k] | i = 1,2,...,l\}$, after transformation.

## IV. Experiment of Irreversibility, Diversity and Revocability

### A. Experimental databases and equipment configuration

The test database of this paper includes PolyU Palmprint Database[27] and CASIA Palmprint Database[28]. The PolyU database consists of 386 palms of 193 individuals, and each palm is collected in two stages, with an interval of two months, about 10 images are collected from each palm at each stage and a total of 7752 palms are obtained. The CASIA palmprint database was established by the Beijing Institute of Automation of the Chinese Academy of Sciences through a collection device without positioning device, the database contains the palmprint images of 310 people, each person collected 8 images of each palm, and finally obtained 5,502 palm print images. The computer used in the experiment is mainly equipped with a processor with a main frequency of 3.4GHz and 8G of memory, and the operating platform is MATLAB.

**B. Analysis of authentication performance**

The authentication performance mainly includes the matching accuracy before the irreversible transformation and the matching accuracy after the irreversible transformation. Here, we select several typical palmprint template protection methods for authentication performance comparison, including BioPhasor[17], BioConvolving[20], RPv[22], RS[23]. The parameters used in the experiment and their specific values are shown in Table 1:

Table 1. List of experimental parameters

| Parameter symbol | Parameter description | Parameter value |
|---|---|---|
| r | The judgment threshold of two directions fusion | 8 |
| n × n | Size of image block | 24 × 24 |
| m | Number of image blocks | 36 |
| l | Number of random Gaussian projection vectors | 420 |
| k | Vector dimension of random Gaussian projection | 50 |

**1) Analysis of matching accuracy before irreversible transformation**

The proposed method was tested on PolyU and CASIA databases, the experimental results are shown in Figure 4, the EER of the proposed method in the two databases was 0.09% and 0.1%, respectively. Compared with other algorithms, as shown in Table 2, the proposed algorithm has the best matching accuracy before irreversible transformation in both databases. Other algorithms represented by BioPhasor generally use Gabor filters or similar methods to extract the direction feature of palmprint, but this paper uses MFRAT of two directions fusion to further improve the accuracy of direction feature extraction, while introducing point feature and direction feature for fusion, experiments prove that the feature extraction method used in this paper has the best matching performance.

Table 2. Comparison of EER (%) before irreversible transformation

| Methods | PolyU | CASIA |
| --- | --- | --- |
| BioPhasor | 0.42 | 0.33 |
| BioConvolving | 0.43 | 0.35 |
| RPv | 0.42 | 0.30 |
| RS | 0.40 | 0.33 |
| **Proposed Method** | **0.09** | **0.10** |

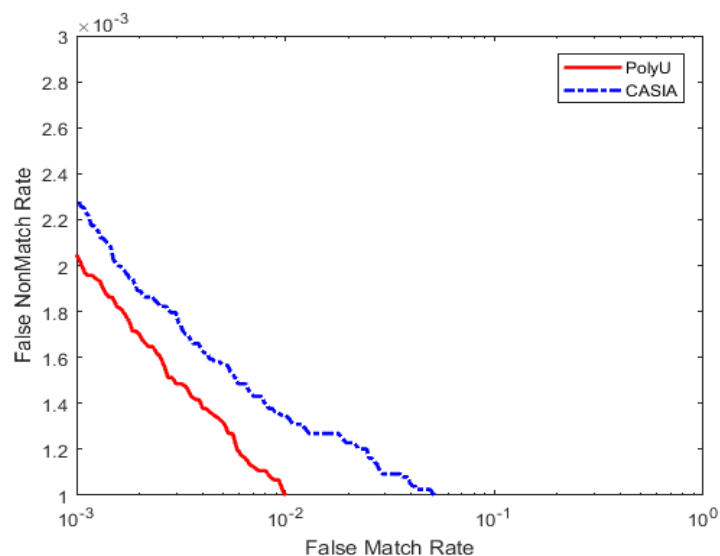

Fig. 4. ROC results of the proposed method before irreversible transformation in the PolyU and CASIA databases

**2) Analysis of matching accuracy after irreversible transformation**

The proposed method was tested on the PolyU and CASIA databases. The experimental results are shown in Figure 5. The EER of the proposed method after irreversible transformation in the two databases was 0.17% and 0.19%, respectively. Compared with other methods, as shown in Table 3, the proposed algorithm has greatly improved the authentication performance in both databases. Compared with BioPhase and other methods, the directional characteristics of the palmprint are mainly extracted by different means and generate the revocable template through irreversible transformation. The proposed method, on the one hand, makes the extracted direction feature more accurate by improving the direction feature extraction algorithm, at the same time, it introduces point features to further enrich the extracted palmprint feature information. On the other hand, the point features are processed in fixed-length and ordered, and the obtained fusion features are more suitable for the IOM irreversible transformation method. The experiment shows that the proposed method is better than the existing best method in matching accuracy after irreversible transformation.

Table 3. EER (%) results after irreversible transformation

| Methods | PolyU | CASIA |
|---|---|---|
| BioPhasor | 1.30 | 1.36 |
| BioConvolving | 5.96 | 2.88 |
| RPv | 0.60 | 0.53 |
| RS | 0.48 | 0.42 |
| **Proposed Method** | **0.17** | **0.19** |

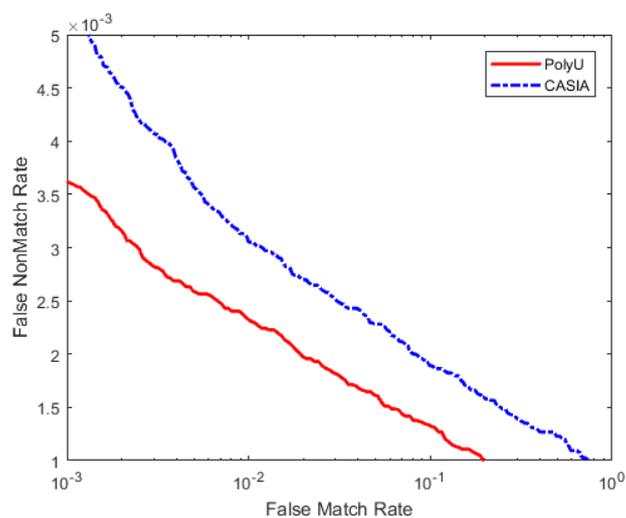

Fig.5. ROC result graph of the proposed method after irreversible transformation in PolyU and CASIA databases

## C. Analysis of accuracy loss

In the template protection method based on transformation, although the security of the revocable template can be guaranteed by irreversible transformation, there is a certain loss of accuracy in palmprint feature matching before and after irreversible transformation. The size of the accuracy loss is mainly determined by the irreversible transformation function, so an excellent irreversible transform function should not only adapt to the extracted palmprint feature, but also keep the matching performance as far as possible without degradation. Compared with other methods, as shown in Table 4, the accuracy loss of the algorithm proposed in this paper is superior, the accuracy loss is the lowest in the CASIA database, and slightly higher than the RS method in the PolyU database. On the whole, the palmprint feature used in this paper is more suitable for the IOM irreversible transformation method, and the final loss of accuracy is lower.

**D. Real-time analysis**

The revocable template generation algorithm proposed in this paper includes the main steps of feature extraction, feature fusion and irreversible transformation, the overall processing time is about 450ms, which is fast and can meet the real-time requirements.

Table 4. Loss of accuracy (%)

| Methods | PolyU | CASIA |
| --- | --- | --- |
| BioPhasor | 0.88 | 1.03 |
| BioConvolving | 5.43 | 2.53 |
| RPv | 0.18 | 0.23 |
| RS | 0.08 | 0.09 |
| Proposed Method | 0.08 | 0.07 |

# V. Conclusion

In this paper, we propose a new composite palmprint feature extraction method for palmprint template protection by fusing direction features and point features. In order to extract more accurate directional features, a direction extraction method based on MFRAT of two directions fusion was proposed. Through the palmprint image is divided into blocks, the representative points are selected in each block based on SURF algorithm, and the fixed-length ordered point features are finally obtained by combining with LBP. The fixed-length ordered direction features are fused with the fixed-length ordered point features to further enrich the extracted palmprint feature information. Finally, IOM irreversible transformation is used to generate a reversible palmprint template for the fused feature.

Compared with other existing works, the experimental results show that our proposed method achieves the best performance on the two databases of PolyU and CASIA. In terms of absolute accuracy loss, the proposed method achieves the best on the CASIA database and comes the second on the PolyU database. In terms of relative accuracy loss, the proposed algorithm is second only to the RS algorithm in both databases.

**Acknowledgement**

We greatly thank the financial supports from the National Cryptography Development Fund (Grant No. MMJJ20170208), the Natural Science Foundation of China (NSFC No. 61876139 and No. 61906149), the Natural Science Basic Research Program of Shaanxi (Program No. 2021JM-136 and No. 2019JM-129).